\documentclass[12pt,twoside]{article}
\usepackage{xspace}
\usepackage{amssymb}
\usepackage{amsthm}
\usepackage{amsmath}
\usepackage{mathrsfs}
\usepackage{fancyvrb}
\usepackage{verbatim}
\usepackage{xcolor}
\usepackage{soul}
\usepackage{url}
\usepackage[
colorlinks,linkcolor={blue},citecolor={blue},urlcolor={red}%
]{hyperref}

\allowdisplaybreaks[4]

\swapnumbers
\newtheorem{theorem}{Theorem}

\theoremstyle{definition}
\newtheorem{remark}[theorem]{Remark}

\newtheorem{problem}[theorem]{Problem}

\DefineVerbatimEnvironment{rlispverb}{Verbatim}{%
framesep=2mm,xleftmargin=\parindent}

\sethlcolor{white}
\newcommand{\cprime}{\/{\mathsurround=0pt$'$}}

\newcommand*{\pd}[2]{\mathchoice{\frac{\partial#1}{\partial#2}}
  {\partial#1/\partial#2}{\partial#1/\partial#2}
  {\partial#1/\partial#2}}

\newcommand*{\fd}[2]{\mathchoice{\frac{\delta#1}{\delta#2}}
  {\delta #1/\delta#2}{\delta#1/\delta#2}{\delta#1/\delta#2}}

\newcommand{\ddx}[1]{\partial_x^{#1}}

\newcommand{\eval}[2][\right]{\relax
  \ifx#1\right\relax \left.\fi#2#1\rvert}

\newcommand{\envert}[2][\right]{\relax
  \ifx#1\right\relax \left\lvert\else#1\lvert\fi#2#1\rvert}

\newcommand{\enVert}[2][\right]{\relax
  \ifx#1\right\relax \left\lVert\else#1\lVert\fi#2#1\rVert}

\let\kappa\varkappa
\let\phi\varphi

\newcommand{\cde}{CDE\xspace}
\newcommand{\cdiff}{CDIFF\xspace}
\newcommand{\cdiffop}{$\mathcal{C}$-differential operator\xspace}
\newcommand{\cdiffops}{$\mathcal{C}$-differential operators\xspace}
\newcommand{\reduce}{Reduce\xspace}

\providecommand{\href}[2]{#2}

\begin{document}
\title{Computing with Hamiltonian operators}

\author{R. Vitolo
  \\
  \footnotesize
Dept.\ of Mathematics and Physics
``E. De Giorgi''
\\
\footnotesize Universit\`a del Salento and INFN, Sezione di Lecce
\\
\footnotesize
via per Arnesano, 73100 Lecce, Italy
\\
\footnotesize
  email: \url{raffaele.vitolo@unisalento.it}
  \\
  \footnotesize
  web: \url{http://poincare.unisalento.it/vitolo}
}

\date{
  \framebox{
\begin{minipage}{7.5cm}
  \begin{center}
  \footnotesize  Published in\\
  \footnotesize
  Computer Physics Communications,\\
  (2019)\\
  \href{https://doi.org/10.1016/j.cpc.2019.05.012}{DOI}
\end{center}
\end{minipage}
}
}

\maketitle

\begin{abstract}
  Hamiltonian operators for partial differential equations are ubiquitous in
  mathematical models of theoretical and applied physics.  In this paper the
  new Reduce package \cde for computations \hl{with} Hamiltonian operators is
  presented. \cde can verify the Hamiltonian properties of skew-adjointness and
  vanishing Schouten bracket for a differential operator, as well as the
  compatibility property of two Hamiltonian operators, and \hl{it can compute}
  the Lie derivative of a Hamiltonian operator with respect to a vector
  field. \hl{More generally, it can compute with} (variational) multivectors,
  or functions on supermanifolds. This can open the way to applications in
  other fields of mathematical or theoretical physics.

  \textbf{Keywords}:  Hamiltonian operators, partial differential equations,
  integrable systems, Schouten bracket, supermanifolds
\end{abstract}

\newpage

\tableofcontents

\section{Introduction}

Hamiltonian operators for partial differential equations (PDEs) are one of the
most important tools in the modern theory of integrable systems
\cite{ablowitz91:_solit,DubrovinKricheverNovikov:InSI,Zakharov:WIsIn}, with
applications ranging from pure to applied mathematics and physics.  Integrable
systems are systems of PDEs for which is it possible to construct classes of
exact solutions in closed form.  A widely accepted characterization of such
systems is that they have infinite sequences of symmetries or conserved
quantities in involution, or \emph{hierarchies}.\footnote{More precisely, they
  commute with respect to certain bracket operations, see~\eqref{eq:19}} Such
quantities play a role in the construction of general solutions \hl{(see}
\cite{ablowitz91:_solit,DubrovinKricheverNovikov:InSI,Zakharov:WIsIn}). It is
not an easy task to show that a certain PDE has a hierarchy.

Consider a system of PDEs of the form
\begin{equation}
  \label{eq:43}
  u_t^i = F^i(x^\lambda,u^j_\sigma),\quad i=1,\ldots, n,
\end{equation}
where $(u^i)$ are dependent variables, $(t,x^\lambda)$ are independent variables
($\lambda=1,\ldots,m$),
\begin{equation}
  \label{eq:44}
  u^i_\sigma = \pd{u^i}{x^{\sigma_1}\cdots\partial x^{\sigma_m}},\quad \sigma
  \in \mathbb{N}^m
\end{equation}
(with $u^i_0=u^i$) and $F^i$ is a smooth function of a finite number of
arguments. The system~\eqref{eq:43} is \emph{Hamiltonian} if there exists a
\hl{linear} matrix differential operator $(A^{ij})$, fulfilling additional
properties, such that
\begin{equation}
  \label{eq:45}
  u_t^i = F^i(x^\lambda,u^j_\sigma) = A^{ij}\fd{H}{u^j},
\end{equation}
where $H$ is a conservation law density, the \emph{Hamiltonian density}, for
the system~\eqref{eq:43}.  The properties that make $A$ a Hamiltonian operator
are: $A$ is skew-adjoint, $A^*=-A$, and the Schouten bracket of $A$ with itself
vanishes, $[A,A]=0$.  We will come back to those properties in
Section~\ref{sec:hamilt-oper-part}.

It can be proved \cite{KrasilshchikVinogradov:SCLDEqMP,KV11} that a Hamiltonian
operator maps conserved quantities into symmetries. But a
Hamiltonian operator alone cannot guarantee the existence of a hierarchy.
The fundamental idea of F. Magri \cite{Magri:SMInHEq} was that if a system of
PDEs is Hamiltonian with respect to \emph{two} Hamiltonian operators $A_1$,
$A_2$, then an infinite sequence of conserved quantities could be generated
through the recursive definition
\begin{equation}
  \label{eq:46}
   A^{ij}_1\fd{H_{k+1}}{u^j} =  A^{ij}_2\fd{H_{k}}{u^j}.
\end{equation}
The sequence is made of commuting conserved quantities with respect to a
certain bracket operation (see~\eqref{eq:19}) if and only if the operators
$A_1$ and $A_2$ are \emph{compatible}: $[A_1,A_2]=0$. Of course, we used the
Schouten bracket between operators.

We can pose the main computational problems of the Hamiltonian formalism for
PDEs.
\begin{enumerate}\label{list:problems}
\item \label{item:4} Given a system of PDEs of the type $u^i_t =
  F^i(x^\lambda,u^i_\sigma)$, find a Hamiltonian operator and a Hamiltonian
  density for the system (direct problem).
\item \label{item:1} Given a Hamiltonian operator $A$, find which integrable
  systems are Hamiltonian with respect to $A$ (inverse problem).
\item \label{item:2} Given a differential operator, check the properties that
  make it Hamiltonian: $A^*=-A$ and $[A,A]=0$.
\item \label{item:2b} Given a Hamiltonian operator $A$, change its coordinates
  in such a way to achieve a certain canonical form (Darboux theorems for
  Hamiltonian operators).
\item \label{item:3} Given two Hamiltonian operators $A_1$, $A_2$, check their
  compatibility: $[A_1,A_2]=0$.
\item \label{item:4b} Given a Hamiltonian operator $A$, compute the set of
  Hamiltonian operators $B$ which are compatible with $A$: $[A,B]=0$.
\item \label{item:5} Given a pair of compatible Hamiltonian operators $A$, $B$,
  find a vector field $\tau$ such that $B=L_\tau A = [\tau,A]$. Here the Lie
  derivative of $A$ with respect to $\tau$ is just the Schouten bracket of the
  vector field with the operator. This problem is of interest in the theory of
  Hamiltonian deformations, see subsection~\ref{sec:comp-lie-deriv}.
\end{enumerate}

While countless papers have been written on the symbolic computation of
symmetries and conserved quantities, the literature on symbolic computations
with Hamiltonian operators is quite scarce. The same situation holds for
publicly available software packages. Possible reasons are that the algorithms
for computing \hl{with} integrability structures, \hl{like Hamiltonian,
  symplectic and recursion operators,} are less standard and more involved with
respect to those for symmetries and conserved quantities. \hl{Another reason
  might be that few experts in integrable systems have the abilities that are
  needed to deal with differential operators in a computer algebra system as
  required for the mathematical problems at hand, and conversely, few experts
  in computer algebra are also familiar with the mathematics of integrable
  systems.}

A recent book on integrability structures \cite{KVV17} tries to fill the above
gaps: it contains, between other material, symbolic computations related to the
direct problem (item~\ref{item:4} above). The software package
\cde\footnote{This is an acronym standing for Calculus on Differential
  Equations}, an official \reduce package \cite{reduce}, is used throughout the
book. \hl{An extensive review of existing software for integrability structures
  can be found in the book.}

\hl{This paper integrates material of}~\cite{KVV17} \hl{in the direction of the
  solution of the above problems 3--7.}  It should be remarked that many
authors developed software for similar purposes without making it available to
the public. That makes their computations much more difficult to
reproduce. There are few software packages with similar capabilities that are
also \hl{in the public domain. We will briefly list them and comment the
  features that are not already considered in the above mentioned review}
\cite[Section 1.3]{KVV17}.

\begin{enumerate}

\item \label{pack1}The Maple package \texttt{jets} \cite{barakat01}
  contains a module for computing the Schouten bracket of local differential
  operators.  \hl{See} \cite[Section 1.3]{KVV17} \hl{for more comments}.
  At the time of writing it seems that the package is no longer available in
  internet, in particular it is not in the website of the author.
\item The Maple package \texttt{Jets} \cite{BMJets}, initially developed by
  Marvan (2003), then also by Baran (2010), has been used for computing
  Hamiltonian operators for particular differential equations
  \cite{BKMV15,BKMV14,BKMV16,BaranMarvan:InWSFC}. However, it does not contain
  any specific feature for integrability structures (like an implementation of
  the Schouten bracket).

\item \label{pack2}The Maple package \texttt{JET} \cite{meshkov02:_tools_pdes}
  can compute with integrability structures, \hl{see} \cite[Section 1.3]{KVV17}
  \hl{for more comments}.

  The limits of the package are the fact that it can only compute with
  operators in one independent variable ($x$) and a partial support of
  computations with \hl{nonlocal} (pseudodifferential) operators $D_x^{-1}$
  (simplification of such terms cannot always be performed). \hl{The recent
    paper}~\cite{CLV19} \hl{provides an algorithm for the
    simplification of such expressions; the authors plan to implement the
    algorithm in computer algebra system (also in} \cde\hl{) in the near
    future.}

\item \label{pack3}The Maple package \texttt{DifferentialGeometry}
  \cite{anderson17:_integ_system_tools}, developed by I. Anderson, is being
  extended by a library of procedures which are devoted to integrable systems
  \hl{(private communication of I. Anderson to the author)}. At the moment the
  \hl{library} is still not available in Maple.

\item The package~\cite{BH10} deserves a mention, in that it is devoted to
  finding recursion operators for symmetries; however, the kind of computations
  that are required are different from those considered here, even if the
  subject is \hl{closely related under the mathematical viewpoint}
  (see~\cite{KVV17} for a deeper discussion).

\item \label{pack4}\hl{In the framework of the algebraic approach to the
    Hamiltonian formalism for PDEs} \cite{BarakatSoleKac:PVAlTHEq} \hl{the
    Mathematica packages \texttt{MasterPVA} and \texttt{WAlg}}
  \cite{casati16:_master_walg} \hl{provide procedures for computing the
    Schouten bracket between local differential operators. The recent
    paper}~\cite{CLV19} \hl{shows that the algebraic approach to Schouten
    brackets leads to the same computations and results as more traditional
    approaches, also for nonlocal operators.}
\end{enumerate}

Basically, the only packages that are ready to be used for computations with
Hamiltonian operators in a set of problems which is wide enough are the above
items \ref{pack2} and \ref{pack4}. The package \texttt{JETS} is limited by the
fact that it can only use operators in one independent variable. The package
\texttt{MasterPVA} \hl{uses an} algebraic formalism which has a completely
different nature \hl{to formalisms which are more common in the field}.
Moreover, both the above packages cannot do the computations of items
\ref{item:2b} and \ref{item:5}, \hl{because the packages lack of capabilities
  to linearize an operator and take its adjoint, and cannot deal with
  multivectors}. Indeed, Hamiltonian operators are bivectors, and computing
with Hamiltonian operators amounts at doing computations for multivectors in
the special case of degree 2, see Section~\ref{sec:hamilt-oper-part}.

The above limits result from reading the manuals; maybe with some extra
programming both packages might include the above capabilities.

The package \cde can compute with operators in an arbitrary number of
variables, it uses a standard formalism or the formalism of supermanifolds
in order to implement the operators and it can do all computations on
Hamiltonian operators that we listed above.

Moreover, \cde is able to deal with Hamiltonian operators as variational
multivectors. \hl{They are dual to variational forms, exactly as ordinary
  multivectors on finite-dimensional manifolds are dual to differential forms.
  Variational forms are endowed by the variational differential (that is the
  analogue of the de Rham differential of forms on manifolds) and constitute
  the so-called \emph{variational complex}} \cite{Vitolo:VS}; \hl{variational
  multivectors are endowed by a bracket operation, the Schouten bracket
  (which is the analogue of the Schouten--Nijenhuis bracket between ordinary
  multivectors)} \cite{IgoninVerbovetskyVitolo:VMBGJS}.

Variational multivectors are implemented in \cde both as differential operators
in several arguments and as superfunctions on supermanifolds. This makes \cde
unique within existing software for computing with Hamiltonian operators, also
in view of future research with multivectors or in the mathematics and physics
of supermanifolds. Indeed, a few packages that can compute with anticommuting
(or Grassmann) variables were written so far
\cite{cheb-terrab95:_grassmann,hartmann89:_grass,peeters06}, but none of them
has an implementation of all the required capabilities for Hamiltonian
operators. At the moment only \cde has an implementation of total derivatives
and variational derivatives that act on functions of even or odd field
variables and their derivatives. \hl{Many calculations that are typical in
  theoretical physics, like computing the Euler--Lagrange equations of a
  Lagrangian defined on a supermanifold, computing symmetries and conservation
  laws for super-differential equations, etc., can be done on a computer by
  means of \cde, with little or no extra programming.} See also
\cite{KrasilshchikKersten:SROpCSDE}, where \cdiff, the `ancestor' of \cde, was
used with a non-automatic implementation of total derivatives for computing
symmetries of super-differential equations; similar tasks were also considered
in~\cite{ayari97:_glie} with a Maple program.

\cde is implemented in \reduce for several reasons: \reduce is free software,
and it was possible to study its internals. As a side outcome of this work,
many undocumented parts of \reduce internals were described in the manual
\cite{NV14}. Then, \reduce has been used for decades with its package \cdiff
for problems which are closely related to those presented in this paper, in
particular by Kersten and the University of Twente research group.
Finally, \cde has an easy interface that makes the above problems easy to
program, with few commands needed to achieve the result.

\hl{Nonlocal Hamiltonian operators are an essential part of integrable systems.
  At the moment,} \cde \hl{can quite easily compute the conditions for a
  nonlocal operator to be Hamiltonian for a given partial differential equation
  (in the sense of mapping conserved quantities into symmetries), see}
\cite{KVV17} \hl{for examples. However,} \cde \hl{lack of capabilities for
  computing Schouten brackets of pseuodifferential operators. In order to fill
  this gap an algorithm was recently developed}~\cite{CLV19}. \hl{A more
  advanced geometric approach is in
  development}~\cite{krasilshchik18:_nonloc_jacob}. \hl{We hope to extend the
  range of} \cde \hl{to nonlocal operators in a near future.}

The structure of the paper is as follows.

In Section~\ref{sec:hamilt-oper-part} we describe multivectors and their
calculus, also in terms of superfunctions. The isomorphism between multivectors
and superfunctions allows to use a very elegant formula for the Schouten
bracket.

In Section~\ref{sec:impl-cdiff-cde} we describe the \cde implementation of
operators and related computations.

In Section~\ref{sec:five-easy-pieces} we show examples of nontrivial
computations performed with \cde. \hl{The examples cover the problems 4--7
  listed above. While the corresponding results have already been published
  elsewhere, the computational methods are a fundamental part of the research
  effort that led to the scientific results, and have never been published
  before. They can be of help for similar computations in other mathematical
  problems. Finally, we present several research-grade problems at the end of
  each example. Hopefully, they will motivate the Reader to using} \cde \hl{in
  his/her scientific research.}

\section{Hamiltonian operators and partial differential
\\  equations}
\label{sec:hamilt-oper-part}

  Let us denote independent variables by $x^\lambda$, $1\leq \lambda\leq m$,
dependent variables by $(u^i)$, $1\leq i\leq n$, and derivatives by
$u^i_\sigma$ (here $\sigma=(\sigma_1,\ldots,\sigma_m)$ is a multiindex
representing derivatives with respect to $x^1$ $\sigma_1$ times, \dots, $x^m$
$\sigma_m$ times).  For each coordinate $x^\lambda$ the \emph{total
  derivative vector field} $\partial_\lambda$ is defined as
\begin{equation}\label{eq:8}
  \partial_\lambda = \pd{}{x^\lambda} + u^i_{\sigma,\lambda}\pd{}{u^i_\sigma},
\end{equation}
(the summation convention holds as usual) where $\sigma,\lambda$ stands for the
multiindex $\sigma+(0,\ldots,1,\ldots,0)$, where $1$ is at the position
$\lambda$. Note that if $\sigma=0$ then $u^i_\sigma = u^i$.

Hamiltonian operators are members of a broader family of differential
operators, namely, they are \emph{variational
multivectors}~\cite{KrasilshchikVinogradov:SCLDEqMP,GelfandDorfman:HOpAlSRT,
Olver:ApLGDEq,IgoninVerbovetskyVitolo:VMBGJS}.
Variational multivectors are specific differential operators in total
derivatives, or \cdiffops (see \cite{KrasilshchikVinogradov:SCLDEqMP} for a
definition).

Let us define functional $m$-forms (or Lagrangians) $\alpha$ as volume forms on
the space of independent variables whose coefficients depend on
$(x^\lambda,u^i_\sigma)$:
\begin{equation}
  \label{eq:5}
  \alpha = \alpha_0(x^\lambda,u^i_\sigma)dx^1\wedge\cdots\wedge dx^m
\end{equation}
The space of functional $m$-forms is denoted by $\bar{\Lambda}^m$.  The kernel
of the variational derivative of local functionals consists of \emph{total
  divergencies}, or functional $m$-forms of the type
$\partial_\lambda(\beta^\lambda)dx^1\wedge\cdots\wedge dx^m$, where $(\beta^\lambda)$,
with $1\leq \lambda\leq m$, is a vector function of $(x^\mu,u^j_\sigma)$. Let us
denote the space of total divergencies by $\bar{d}\bar{\Lambda}^{m-1}$. We
define the space of \emph{local functionals} as $\bar{H}^m =
\bar{\Lambda}^m/\bar{d}\bar{\Lambda}^{m-1}$.

A variational multivector is defined to be a skew-symmetric
\cdiffop with values in local functionals.

The expression of a variational multivector is
\begin{equation}
  \label{eq:20}
  \Delta(\psi^1,\dots,\psi^h) = 
  [a{}^{(\sigma_1 i_1)}{}^{\cdots}{}^{(\sigma_h i_h)}
 \partial_{\sigma_1} \psi^1_{i_1}\cdots \partial_{\sigma_h}\psi^h_{i_h}].
\end{equation}
Here, the arguments of $\Delta$ are vector-valued functions of the type
$\psi = \big(\psi_i(x^\lambda,u^j_\sigma)\big)$, and the coefficients
$a{}^{(\sigma_1 i_1)}{}^{\cdots}{}^{(\sigma_h i_h)}$ are functions of
$(x^\lambda,u^j_\sigma)$ which are skew-symmetric with respect to the interchange
of pairs $(\sigma_k i_k)$ and $(\sigma_h i_h)$.  The value of $\Delta$,
i.e., the right-hand side of \eqref{eq:20}, is an equivalence class up to total
divergencies. Note that we will omit $dx^1\wedge\cdots\wedge dx^m$ for the sake
of brevity.

A variational multivector can be uniquely represented by taking its formal
adjoint in one of its arguments (say the last one; see
\cite{KrasilshchikVinogradov:SCLDEqMP} for a geometric definition):
\begin{equation}
  \label{eq:1}
  \Delta^*(\psi^1,\dots,\psi^{h-1})^i =
  (-1)^{|\sigma_h|}\partial_{\sigma_h}(a{}^{(\sigma_1 i_1)}{}^{\cdots}{}^{(\sigma_h i)}
 \partial_{\sigma_1} \psi^1_{i_1}\cdots \partial_{\sigma_{h-1}}\psi^{h-1})
\end{equation}
Indeed, it is easy to realize that the above expression is equal to the
previous expression (up to a total divergence) and it is divergence-free, i.e.,
no expression of order zero in one of the arguments can have the form of a
total divergence.

The above representation defines an isomorphism between the space of
variational $h$-vectors and the space of vector-valued $h$-\cdiffops of the form
\begin{equation}
  \label{eq:2}
  \square(\psi^1,\dots,\psi^{h})^j =
  b{}^j{}^{(\sigma_1 i_1)}{}^{\cdots}{}^{(\sigma_{h} i_{h})}
  \partial_{\sigma_1} \psi^1_{i_1}\cdots \partial_{\sigma_{h}}\psi^{h}_{i_{h}}
\end{equation}
which are both skew-symmetric with respect to the exchange of arguments and
skew-adjoint with respect to each argument
\cite{IgoninVerbovetskyVitolo:VMBGJS}.

The calculus of variational multivectors \hl{consists of obvious operations},
like sums and compositions, and of the variational Schouten bracket (or just
Schouten bracket for short). The Schouten bracket for variational multivectors
was first formulated in wide generality in \cite{GelfandDorfman:HOpAlSRT} (see
also
\cite{IgoninVerbovetskyVitolo:VMBGJS,KerstenKrasilshchikVerbovetsky:HOpC}).
\hl{Its expression is not implemented in} \cde \hl{in the language of
  differential operators.}

There is another way to express the bracket which is much more elegant and
compact \cite{Getzler:DTHOpFCV} (see also
\cite{IgoninVerbovetskyVitolo:VMBGJS,IgoninVerbovetskyVitolo:FLVDOp,
  KerstenKrasilshchikVerbovetsky:HOpC}). Let us consider a vector of new
dependent variables $(p_i)$, $1\leq i \leq n$. We will assume that the new
variables (and hence their derivatives) are Grassmann, or anticommuting,
variables, so that we will be dealing with the jet of a superbundle with
coordinates $(x^\lambda,u^i_\sigma,p_{j\tau})$. Then, there is an isomorphism
between the space of skew-symmetric vector-valued \cdiffops~\eqref{eq:2} and
the space of vector-valued superfunctions
\begin{equation}
  \label{eq:17}
  F^j = b{}^j{}^{(\sigma_1 i_1)}{}^{\cdots}{}^{(\sigma_h i_h)}
        p_{i_1\sigma_1} \cdots p_{i_h\sigma_h}.
\end{equation}
In coordinates, the isomorphism is given by
\begin{equation}
  \label{eq:223}
  b{}^j{}^{(\sigma_1 i_1)}{}^{\cdots}{}^{(\sigma_h i_h)}
 \partial_{\sigma_1} \psi^1_{i_1}\cdots \partial_{\sigma_h}\psi^h_{i_h}
    \quad\longrightarrow\quad
     b{}^j{}^{(\sigma_1 i_1)}{}^{\cdots}{}^{(\sigma_h i_h)}
        p_{i_1\sigma_1} \cdots p_{i_h\sigma_h}
\end{equation}
Of course, if the skew-symmetric \cdiffop is a variational multivector, then
the corresponding superfunction is defined up to total divergencies of
vector-superfunctions.

Using the above formalism, if two variational multivectors are represented by
the scalar superfunctions $F$ and $H$ (up to total divergencies), then we have
the following formula for the Schouten bracket:
\begin{equation}
  \label{eq:4}
  [F,H] = \left[
    \fd{H}{u^i} \fd{F}{p_i} - (-1)^{(F+1)(H+1)}\fd{F}{u^i} \fd{H}{p_i}
    \right]
\end{equation}
Here the letters $F$, $H$ at the exponent of $(-1)$ mean the Grassmann parity
of the corresponding superfunction, and $\fd{F}{u^i}$, $\fd{F}{p_i}$,\dots
denote the variational derivatives
\begin{equation}
  \label{eq:16}
  \fd{F}{u^i} = (-1)^{|\sigma|}\partial_\sigma\left(\pd{F}{u^i_\sigma}\right),\quad
  \fd{F}{p_i} = (-1)^{|\sigma|}\partial_\sigma\left(\pd{F}{p_{i\sigma}}\right).
\end{equation}
Note that the derivatives with respect to odd coordinates are odd derivatives,
and total derivatives~\eqref{eq:8} are extended to odd variables as
\begin{equation}
  \label{eq:12}
  \partial_\lambda = \pd{}{x^\lambda} + u^i_{\sigma,\lambda}\pd{}{u^i_\sigma}
  + p_{i\sigma,\lambda}\pd{}{p_{i\sigma}}.
\end{equation}
The square brackets on the right-hand side of~\eqref{eq:4} mean that we are in
an equivalence class, or that the expression in the bracket is considered up to
total divergencies of superfunctions.

The above formula is undoubtedly the clearest and most elegant expression
of the Schouten bracket. But its clarity has a price to pay: one should
be able to compute with Grassmann variables in order to use it. We will see that
\cde implements the required capabilities.

Hamiltonian operators are (matrix) differential operators in total derivatives
acting on vector-valued functions $\psi =\big(\psi_i(x^\lambda,u^j_\sigma)\big)$
\begin{equation}
  \label{eq:7}
  A(\psi) = A^{ij} \psi_j = a^{i(\sigma j)}\partial_\sigma \psi_j,
\end{equation}
where $a^{i(\sigma j)}$ are functions of $(x^\lambda,u^k_\sigma)$.

The operator $A$ is requested to fulfill two properties:
\begin{itemize}
\item $A$ is formally skew-adjoint: $A^* = -A$, where
  \begin{equation}
  (A^*)^{ij}\psi_j = (-1)^{|\sigma|}\partial_\sigma\left(\psi_ja^{j(\sigma
      i)}\right)\label{eq:9}
\end{equation}
This means that $A$ is a variational bivector represented as~\eqref{eq:2};

\item The \emph{Schouten bracket} of the operator with itself is zero:
  \begin{equation}
  [A,A]=0.\label{eq:22}
\end{equation}

\end{itemize}
In order to compute the Schouten bracket of $A$, we should first of all
represent $A$ as a superfunction. In order to use the isomorphism \eqref{eq:223}
we need to write $A$ in the form \eqref{eq:20}. This just means that the
isomorphism takes the form
\begin{equation}
  \label{eq:13}
  a^{i(\sigma j)}\partial_\sigma \psi^1_j\psi^2_i
  \quad\longrightarrow\quad
   a^{i(\sigma j)}p_{j\sigma}p_i,
\end{equation}
where the product between $p_{j\sigma}p_i$ is the Grassmann product.
Then, as the Grassmann parity of $A$ is $2$, we have
\begin{equation}
  \label{eq:14}
  [A,A] = 2\left[
    \fd{A}{u^i} \fd{A}{p_i}
    \right]
  \end{equation}
where $A=a^{i(\sigma j)}p_{j\sigma}p_i$. If we wish to check that the Schouten
bracket is zero, we can apply the Euler--Lagrange operator to the expression
in the bracket to verify that it is zero.

A partial differential equation in evolutionary form
$u^i_t = f^i(x^\lambda,u^j,u^j_\sigma)$ is said to be \emph{Hamiltonian} with
respect to a Hamiltonian differential operator $A = (A^{ij})$ if it exists a
local functional $H = [h]$, where $h=h(x^\lambda,u^i_\sigma)$, such that
\begin{equation}
  \label{eq:3}
  u^i_t = f^i(x^\lambda,u^j,u^j_\sigma) = A^{ij}\left(\fd{h}{u^j}\right).
\end{equation}
It can be proved that a Hamiltonian operator
satisfies the following equation:
\begin{equation}
  \label{eq:6}
  \ell_F\circ A = - A\circ \ell^*_F,
\end{equation}
where $\ell_F$ is the Fr\'echet derivative, or linearization, of the function
$F=u^i_t - f^i$ that defines the equation, and $\ell_F^*$ is its formal
adjoint:
\begin{equation}
  \label{eq:15}
  \ell_F(\varphi) = \pd{F^k}{u^i_\sigma}\partial_\sigma\varphi^i,
  \quad
  \ell^*_F(\psi) = (-1)^{|\sigma|}\partial_\sigma\left(
    \pd{F^k}{u^i_\sigma}\psi_k
  \right),
\end{equation}
We observe that the kernel of $\ell_F$ \hl{consists} of (generalized, or higher)
\emph{symmetries} of the differential equation \eqref{eq:3}
\cite{KrasilshchikVinogradov:SCLDEqMP,Olver:ApLGDEq}. A conservation law
$\omega= h dt + k dx$ for an evolutionary equation~\eqref{eq:3} is defined up
to trivial quantities of the form $\alpha = \partial_t f dt + \partial_x f dx$,
and is uniquely represented by its \emph{generating function}, or
\emph{characteristic}, $\psi=(\fd{h}{u^i})$. The kernel of $\ell^*_F$ contains
generating functions, or characteristic vectors, of conserved quantities
\cite{KrasilshchikVinogradov:SCLDEqMP,Dorfman:DSInNEvEq,Olver:ApLGDEq}.  Hence,
\eqref{eq:6} implies that a Hamiltonian operator maps conserved quantities into
symmetries.

An extension of this notion for non-evolutionary equations is also available
\cite{KerstenKrasilshchikVerbovetskyVitolo:HSGP}, but will not be considered
here.

The \emph{Poisson bracket} of local functionals $H=[h]$ and $F=[f]$ is
\begin{equation}
  \label{eq:19}
  \{H,F\}_A = \left[\fd{h}{u^i}A^{ij}\fd{f}{u^j}\right].
\end{equation}
The Poisson bracket fulfills the properties
\begin{gather}
  \label{eq:21}
  \{H,F\}_A = - \{F,H\}_A
  \\
  \label{eq:11}
  \{\{H,F\}_A,G\}_A + \{\{G,H\}_A,F\}_A + \{\{F,G\}_A,H\}_A = 0
\end{gather}
and it endows the space of functionals with the structure of a Lie algebra.  It
can be proved that the above properties are direct consequences of the
properties~\eqref{eq:9} and \eqref{eq:22}, respectively; and conversely, if the
bracket defined by a \cdiffop $A$ as in \eqref{eq:19} fulfills \eqref{eq:21}
and \eqref{eq:11} then $A$ is a Hamiltonian operator. Of course, the bracket
can be computed between generating functions of conserved quantities.  Usually,
a partial differential equation is said to be \emph{integrable} if it possesses
a sequence of conserved quantities which are in involution with respect to a
Poisson bracket.

The integrability of a partial differential equation follows when the equation
admits two Hamiltonian formulations~\eqref{eq:3} for two distinct Hamiltonian
operators $A_1$, $A_2$ and respective densities $H_1$, $H_2$; the operators are
required to be \emph{compatible}, or that their Schouten bracket vanish:
$[A_1,A_2]=0$. In this case, the celebrated theorem by Magri
\cite{Magri:SMInHEq} states that \hl{the conserved quantities generated
by the recurrence formula}
\begin{equation}
  \label{eq:10}
  A_1(\psi_{i+1}) = A_2(\psi_i)
\end{equation}
\hl{are in involution with respect to the Poisson
brackets defined by both operators:} $\{\psi_i,\psi_{i+1}\}_{A_1} =
\{\psi_i,\psi_{i+1}\}_{A_2}=0$, \hl{$i=0,1,2,$\dots}

\section{Computing with \cdiffops and superfunctions}
\label{sec:impl-cdiff-cde}

We describe the implementation of \cdiffops and superfunctions in \cde and
related operations. The exposition differs from that of~\cite{KVV17} in that it
is aimed at computing with Hamiltonian operators.

Standard \reduce concepts and syntax will not be discussed here; we invite the
interested reader to have a look at \reduce's official website
\cite{reduce}. \hl{More details on} \cde, \hl{including a list of all
  commands, can be found in} \cite{KVV17}.

\subsection{\cde and total derivatives}
\label{sec:cde-jet-space-1}

Starting from independent, dependent and Grassmann, or odd, variables \cde
automatically creates the list of derivative coordinates (as symbols).  The
basic properties of odd variables in \cde, namely the odd product, the odd
derivatives and the definition of a supervectorfield, are provided by the
package \cdiff (now officially in the \reduce distribution), \hl{written} by
Gragert, Kersten, Post and Roelofs from the University of Twente
(Netherlands), which is automatically loaded by \cde.

In a \reduce terminal, load the \cde package by the command
\texttt{load\_package cde;}. Then \hl{a} supermanifold is created by the
command
\begin{rlispverb}
  cde({indep_var,dep_var,odd_var,total_order},{});
\end{rlispverb}
Here, the first list contains the names of the lists of independent variables,
dependent variables, odd variables and the maximal order of derivative
coordinates. For example, we might have
\begin{rlispverb}
  indep_var:={t,x};  dep_var:={u,v};
  odd_var:={p,q};  total_order:=10;
\end{rlispverb}
The empty list at the second argument of \texttt{cde} is used for the
restriction of total derivatives on differential equations (which we will not
use in this paper). After the above command the system has distinct lists of
even derivatives $(u^i_\sigma)$ and odd derivatives $(p_{j\tau})$ (up to the
order \texttt{total\_order}) in the form
\begin{rlispverb}
  u_x,v_t,u_t2x,v_tx,v_2t3x,...,p_t2x,...
\end{rlispverb}
Note that the symbol \texttt{v\_xt} does not exist in \cde (with the above
data) for performance reasons. Odd variables can appear in anticommuting
products; this is represented as
\begin{rlispverb}
  ext(p,p_2tx),ext(p_x,q_t,q_t2x),...
\end{rlispverb}
where \texttt{p} and \texttt{ext(p)} are just the same, while
\texttt{ext(p\_2tx,p)} is not automatically expanded to \texttt{-ext(p,p\_2tx)}
for performance reasons; only one of the two symbols is understood by the
system. In order to get the correct representation the function
\begin{rlispverb}
  odd_product(phi,psi);
\end{rlispverb}
should be used for the odd product of two superfunctions.  The derivative of an
expression \texttt{phi} with respect to an odd variable \texttt{p} is achieved
by \texttt{df\_odd(phi,p);}.

\cde also defines total derivatives~\eqref{eq:12} on the given supermanifold,
truncated at the order \texttt{total\_order}. Total derivatives are
distinguished supervector fields; they are defined through a \cdiff function
using data that is constructed by \cde. Basically, for each derivative symbol
(like \texttt{u\_2tx}) in the system the coefficient of the corresponding
derivative $\pd{}{u_{2tx}}$ is computed by increasing the multiindex by one in
the direction of the corresponding \hl{independent} variable, and the results
are summed up to make~\eqref{eq:12}. The total derivative of a superfunction
\texttt{phi} is invoked by \texttt{td(phi,x);} or \texttt{td(phi,t,x,2);}, for
example.  The syntax closely follows \reduce syntax for standard derivatives
\texttt{df}; the above expressions translate to $\partial_x\varphi$ and
$\partial_t\partial_x^2\varphi$, respectively.

\hl{Each time that a coefficient in a total derivative has order higher than
  \texttt{total\_order}} it is replaced by the identifier
\texttt{letop}\footnote{In Dutch `let op' means `pay attention'}, which is a
function that depends on \hl{all} independent variables. \hl{After evaluating
  total derivatives, the result is scanned} for the presence of \texttt{letop}.
\hl{If that is the case the program stops with an error message} and the
computation must be repeated with a higher \hl{value of} \texttt{total\_order}.
\hl{If needed,} \cde \hl{programs can be run by a Linux shell script,
  \texttt{rrr.sh}, which is included in the package; the script will re-run the
  program with a higher jet order if a jet order error is met.}

The computation of total derivatives \hl{can lead to huge expressions}. \hl{It
  is possible to disable the expansion of total derivatives} by the command
\texttt{noexpand\_td();} (and re-enable it by \texttt{expand\_td();}).

The command
\begin{rlispverb}
pvar_df(par,expr,dvar)
\end{rlispverb}
computes the variational derivative $\fd{F}{w^i}$, where $F$ is \texttt{expr},
$w^i$ is the dependent variable \texttt{dvar} and \texttt{par} is the parity of
the dependent variable, $0$ if even or $1$ if odd. The command
\texttt{euler\_df(expr)} computes all even and odd variational derivatives and
returns their values in a list of two lists.

\subsection{\cde and \cdiffops}
\label{sec:cde-cdiffops}

A vector valued \cdiffop~\eqref{eq:2} must be declared in \cde as follows:
\begin{rlispverb}
  mk_cdiffop(opname,num_arg,length_arg,length_target);
\end{rlispverb}
where
\begin{itemize}
\item \texttt{opname} is the name of the operator ($\square$ in \eqref{eq:2});
\item \texttt{num\_arg} is the number of arguments ($h$ in \eqref{eq:2});
\item \texttt{length\_arg} is the list of lengths of the arguments, e.g., in
  \eqref{eq:2} one needs a list of $h$ items \texttt{\{k\_1,\dots,k\_h\}},
  each corresponding to number of components of the vector functions
  $\psi^j_{i_j}$ to which the operator is applied. In the calculations of this
  paper we will only use one argument;
\item \texttt{length\_target} is the number of components of the image vector
  function (the range of the index $j$ in \eqref{eq:2}).
\end{itemize}
The above parameters of the operator \texttt{opname} are saved in the
\emph{property list} of the identifier \texttt{opname} \hl{(more comments can
  be found in} \cite{KVV17}).  The value of one component of the operator
$\square$ on the arguments $\psi^1$,\dots, $\psi^{h}$ is
\begin{rlispverb}
  opname(j,i1,...,ih,psi1,...,psih);
\end{rlispverb}

A vector-valued superfunction~\eqref{eq:17} must be declared in \cde as
follows:
\begin{rlispverb}
mk_superfun(sfname,deg,length_target);
\end{rlispverb}
where
\begin{itemize}
\item \texttt{sfname} is the name of the superfunction ($F$ in \eqref{eq:17});
\item \texttt{deg} is the degree of the superfunction, e.g., $h$ in
  \eqref{eq:17};
\item \texttt{length\_target} is the number of components of the image vector
  (the range of the index $j$ in \eqref{eq:17}).
\end{itemize}
The syntax for one component of the superfunction \texttt{sfname} is
\begin{rlispverb}
  sfname(j);
\end{rlispverb}

\cde \hl{can convert} \cdiffops with one
argument \hl{into} superfunctions of degree $1$ \hl{and back}:
\begin{itemize}
\item \texttt{conv\_cdiff2superfun(cdop,superfun)}
\item \texttt{conv\_superfun2cdiff(superfun,cdop)}
\end{itemize}
The translation from an operator to a superfunction is easy: it is enough to
evaluate all components of the operator on all odd dependent variables.  The
other direction is more complicated: coefficients $C_i^{j \sigma}$ of odd
variables $p_{j\sigma}$ have to be collected from each component $i$ of the
superfunction, then a Reduce operator \texttt{cdop(i,j,psi)} is defined by the
sum (with respect to $\sigma$) $C_i^{j\sigma}\partial_\sigma \psi$. Here the
technical difficulty is defining an operator inside a Reduce procedure,
\texttt{psi} being just a formal parameter and not an expression. Indeed, as a
data structure, a superfunction is much more easy to handle than a Reduce
operator.

The linearization of a purely even vector-function and
the adjoint of a \cdiffop with one argument are computed by means of the
isomorphism~\eqref{eq:223}. More precisely, given a function
$F=F(x^\lambda,u^i_\sigma)$ the following superfunctions can be easily computed:
\begin{equation}
  \label{eq:18}
  \ell_F((p_j))^k = \sum_{i,\sigma} \pd{F^k}{u^i_\sigma}p_{i_\sigma}
  \quad
  \ell^*_F((p_j))_i =
  (-1)^{|\sigma|}\partial_\sigma\left(\pd{F^k}{u^i_\sigma}p_{k_\sigma}\right)
\end{equation}
and then translated back into \cdiffops using the above conversion utilities.

In \cde a vector function must be introduced as a list of scalar functions
\begin{rlispverb}
fun:={fun1,fun2,...};
\end{rlispverb}
Then its linearization is achieved by the left formula~\eqref{eq:18}
implemented in the command
\begin{rlispverb}
ell_function(fun,lfun);
\end{rlispverb}
where \texttt{lfun} is automatically declared as a \cdiffop with the appropriate
parameters. Moreover, the above command creates a superfunction
\texttt{lfun\_sf} that corresponds to the \cdiffop \texttt{lfun}.

The command
\begin{rlispverb}
adjoint_cdiffop(lfun,lfun_star);
\end{rlispverb}
computes the adjoint \texttt{lfun\_star} of \texttt{lfun} by the formula on the
right of~\eqref{eq:18} and introduces an equivalent superfunction whose
identifier has the suffix \texttt{\_sf}: \texttt{lfun\_star\_sf}.

\section{Five easy pieces}
\label{sec:five-easy-pieces}

Elementary computations with Hamiltonian operators can be found in the \cde
section in Reduce's manual \cite{reduce}. Here, we would like to show the
\hl{solution of the computational problems that we listed in the Introduction
  (p.}\ \pageref{list:problems}, \hl{items 4--7) for nontrivial examples.} The
  examples have been published in separate research papers in recent years.

While results are almost all known, a detailed description of the corresponding
computations appears here for the first time. \hl{The design of effective
  software for the solution of the mathematical problems that have been
  considered so far was is a fundamental and nontrivial part of the research
  activity. It deserves a separate exposition} and it might be of interest when
trying to solve similar problems. All the examples are provided as \reduce
program files in the website of the author
\url{http://poincare.unisalento.it/vitolo}.

To the authors' knowledge, no other computer algebra computations of this type
and level have been published with all details before.

\hl{At the end of each example a research problem is presented in order to
  motivate the Reader to use} \cde \hl{in his/her research.}

\subsection{Warm-up:  tedious large-scale computations}
\label{sec:warmup:-large-scale}

\hl{The associativity equation, or Witten--Dijkgraaf--Verlinde--Verlinde (WDVV)
  equation was derived in the context of $2D$-topological field theory. It
  consists in an overdetermined system of PDEs with one dependent variable and
  $N$ independent variables. Nowadays its significance is mostly
  mathematical. For example, its solutions yield, under mild hypotheses,
  bi-Hamiltonian systems. See} \cite{D96} \hl{for a mathematical introduction
  to the equation.}

In \cite{FGMN97} a bi-Hamiltonian formulation of the WDVV equation in the
simplest case $N=3$ was found. It was only in recent times that such a result
was extended to the case $N=4$, also thanks to \cde \cite{PV15}. Here we will
describe the role of \cde in \cite{PV15}.

In \cite{ferapontov96:_hamil} the WDVV equation was rewritten, in the case
$N=4$ (see \cite{D96} for details), as a pair of commuting hydrodynamic-type
systems
\begin{equation}\label{eq:23}
  \begin{array}{l}
    a^{1}_y = a^{2}_x,
    \\
    a^{2}_y = a^{4}_x
    \\
    a^{3}_y = a^{5}_x
    \\
    a^4_y = R_x
    \\
    a^5_y= P_x
    \\
    a^{6}_y = S_x
  \end{array}
  \quad\text{and}\quad
  \begin{array}{l}
    a^{1}_z = a^{3}_x
    \\
    a^{2}_z = a^{5}_x
    \\
    a^{3}_z = a^{6}_x
    \\
    a^{4}_z = P_x,
    \\
    a^{5}_z = S_x,
    \\
    a^{6}_z = Q_x
  \end{array}
\end{equation}
where%
\begin{gather*}
P=\frac{a^{3}a^{4}+a^{6}}{a^{1}},\text{ \ }R=\frac{2a^{5}+a^{2}a^{4}}{a^{1}},%
\text{ \ }S=\frac{2a^{3}a^{5}-a^{2}a^{6}}{a^{1}},
\\
Q=(a^{5})^{2}-a^{4}a^{6}+\frac{%
(a^{3})^{2}a^{4}+a^{3}a^{6}-2a^{2}a^{3}a^{5}+(a^{2})^{2}a^{6}}{a^{1}}.
\end{gather*}

In \cite{ferapontov96:_hamil} (see also \cite{mokhov98:_sympl_poiss}) it was
proved that the above two systems~\eqref{eq:23} admit a first-order
Dubrovin--Novikov Hamiltonian operator. This is an operator of the type
\begin{equation}
  \label{eq:27}
  A_1^{ij} = h^{ij}\ddx{} + \Gamma^{ij}_k u^k_x.
\end{equation}
The operator is homogeneous (of degree $1$) with respect to a grading which is
given by $x$-derivatives~\cite{DN83}. This implies that the form of the
operator is invariant with respect to coordinate transformations of type
$\tilde{u}^i = \tilde{u}^i(u^j)$. As a consequence, the coefficients transform
as geometric objects: for example, $h^{ij}$ transforms as a contravariant
$2$-tensor. It can be proved that the Hamiltonian property of the operator is
equivalent (here and in what follows $\det(h^{ij})\neq 0$) to the fact that
$(h_{ij})=(h^{ij})^{-1}$ is a flat pseudo-Riemannian metric and
$\Gamma^{j}_{hk}= - h_{hi}\Gamma^{ij}_k$ are its Christoffel symbols.

In our example, the first-order Dubrovin--Novikov Hamiltonian operator
for~\eqref{eq:23} is
\begin{equation}
A_{1}^{ij}=%
\begin{pmatrix}
0 & 0 & 0 & -1 & 0 & 0 \\ 
0 & -1 & 0 & 0 & 0 & 0 \\ 
a^{1} & a^{2} & a^{3} & a^{4} & a^{5} & a^{6} \\ 
-1 & 0 & 0 & 0 & 0 & 0 \\ 
a^{2} & a^{4} & a^{5} & R & P & S \\ 
2a^{3} & 2a^{5} & 2a^{6} & 2P & 2S & 2Q%
\end{pmatrix}%
\ddx{} + \ddx{}
\begin{pmatrix}
0 & 0 & a^{1} & -1 & a^{2} & 2a^{3} \\ 
0 & -1 & a^{2} & 0 & a^{4} & 2a^{5} \\ 
0 & 0 & a^{3} & 0 & a^{5} & 2a^{6} \\ 
-1 & 0 & a^{4} & 0 & R & 2P \\ 
0 & 0 & a^{5} & 0 & P & 2S \\ 
0 & 0 & a^{6} & 0 & S & 2Q%
\end{pmatrix}.  \label{eq:32}
\end{equation}%
In~\cite{PV15} it was proved that the above two systems~\eqref{eq:23} are
indeed bi-Hamiltonian, as they also admit the compatible Dubrovin--Novikov type
third-order Hamiltonian operator \cite{DubrovinNovikov:PBHT} (see
subsection~\ref{sec:find-darb-coord})
\begin{equation}
  \label{eq:24}
  A_2^{ij} = \ddx{}(g^{ij}\ddx{} + c^{ij}_ka^k_x)\ddx{},
\end{equation}
where we assume that $\det(g^{ij})\neq 0$, and if we set
$(g_{ij})=(g^{ij})^{-1}$, then
\begin{equation}
  \label{eq:25}
  g_{ik}(\mathbf{a})=%
\begin{pmatrix}
(a^{4})^{2} & -2a^{5} & 2a^{4} & -(a^{1}a^{4}+a^{3}) & a^{2} & 1 \\ 
-2a^{5} & -2a^{3} & a^{2} & 0 & a^{1} & 0 \\ 
2a^{4} & a^{2} & 2 & -a^{1} & 0 & 0 \\ 
-(a^{1}a^{4}+a^{3}) & 0 & -a^{1} & (a^{1})^{2} & 0 & 0 \\ 
a^{2} & a^{1} & 0 & 0 & 0 & 0 \\ 
1 & 0 & 0 & 0 & 0 & 0%
\end{pmatrix}.%
\end{equation}
The coefficients $c^{ij}_k$ are given by the formula \cite{FPV14}
\begin{equation}
  c_{skm}=\frac{1}{3}(g_{sm,k}-g_{sk,m}),  \label{eq:195}
\end{equation}
where $c_{ijk}=g_{iq}g_{jp}c_{k}^{pq}$. The condition $[A_2,A_2]=0$ is ensured
by the properties \cite{FPV14,potemin97:_poiss}
\begin{align}
& g_{mk,s}+g_{ks,m}+g_{ms,k}=0,  \label{eq:2011} \\
& c_{msk,l}=-g^{pq}c_{pml}c_{qsk}.  \label{eq:2111}
\end{align}%

In \cite{PV15} \cde \hl{was used to check that} $[A_1,A_2]=0$. Indeed, for few
types of Dubrovin--Novikov operators \hl{(like first, second and third-order
  operators)} it is known that the vanishing of the Schouten bracket is
equivalent to some tensorial conditions on the coefficients of the
operators. However, at the moment of writing, tensorial conditions of
compatibility of a first order and a third-order Dubrovin-Novikov operator are
not known. Hence, \hl{the only way to check that $[A_1,A_2]=0$ is a direct
  computation}. This is a very long and tedious task which is not instructive
at all: a typical computation for a machine.

Here we describe the implementation of the calculation of $[A_1,A_2]=0$.  The
program file is \texttt{w6c\_biham} and can be found at the web page of the
author~\url{http://poincare.unisalento.it/vitolo}.

In \reduce, load the package \cde, then declare the input variables and call
\texttt{cde}:
\begin{rlispverb}
indep_var:={x};
dep_var:={a,b,c,d,ee,f};
odd_var:={p,q,r,s,tt,u};
total_order:=10;
cde({indep_var,dep_var,odd_var,total_order},{});
\end{rlispverb}


Then we define two matrices whose entries are: the metric of the first-order
operator (in upper indices) \texttt{hu1(i,j)}, and the metric of the
third-order operator (in lower indices) \texttt{gl3(i,j)}.  We define two
operators, \texttt{gamma\_hi\_con(i,j)} that contains the expression
$\Gamma^{ij}_k u^k_x$ and \texttt{c\_hi\_con(i,j)} that contains the expression
$c^{ij}_k u^k_x$. The operator $A_1$ is defined as
\begin{rlispverb}
mk_cdiffop(aa1,1,{6},6);
for all i,j,psi let aa1(i,j,psi)=
 hu1(i,j)*td(psi,x)+gamma_hi_con(i,j)*psi;
\end{rlispverb}
and the operator $A_2$ is defined as:
\begin{rlispverb}
mk_cdiffop(aa2,1,{6},6);
for all i,j,psi let aa2(i,j,psi) =
td(
gu3(i,j)*td(psi,x,2)+c_hi_con(i,j)*td(psi,x)
,x);
\end{rlispverb}
We convert them into superfunctions, according with \eqref{eq:223}
\begin{rlispverb}
conv_cdiff2superfun(aa1,aa1_sf);
conv_cdiff2superfun(aa2,aa2_sf);
\end{rlispverb}
We take their adjoint by \texttt{cde} and make the following simple
skew-adjointness test:
\begin{rlispverb}
adjoint_cdiffop(aa1,aa1_star);
for i:=1:length(dep_var) do
  if aa1_sf(i) + aa1_star_sf(i) neq 0 then
    write "Warning: non-skew-adjoint operator!";
\end{rlispverb}
and analogously for \texttt{aa2}. Then, we convert the operators into
bivectors, according with~\eqref{eq:2}
\begin{rlispverb}
conv_genfun2biv(aa1_sf,biv1);
conv_genfun2biv(aa2_sf,biv2);
\end{rlispverb}
Finally, we should check whether the Schouten bracket of the two operators is
zero. We can even compute all possible Schouten brackets, to check the
Hamiltonian property of the two operators:
\begin{rlispverb}
iszero_schouten_bracket(biv1,biv1,thr11b);
iszero_schouten_bracket(biv1,biv2,thr12b);
iszero_schouten_bracket(biv2,biv2,thr22b);
\end{rlispverb}
the results are lists of zeros, and the computation takes a negligible time on
a contemporary laptop of average computing power.

\begin{problem}
  \hl{When forming WDVV equations an essential parameter is the number $N$ of
    independent variables. It is an opinion of the author that WDVV equations
    might have a bi-Hamiltonian formalism for an arbitrary value of $N$.  It is
    a recent discovery that the same holds for another system of PDEs that is
    of fundamental importance in the geometric theory of integrable systems,
    the oriented associativity equation, in the case $N=3$}
  \cite{casati19:_hamil}.

\hl{The Readers might wish to try to find a bi-Hamiltonian formalism for the
  WDVV system in the easiest unknown case $N=5$, to corroborate (or negate!)
  the above conjecture. The recommended literature for this problem is}
  \cite{FGMN97,FPV17:_system_cl,PV15}.
\end{problem}

\subsection{Finding Darboux coordinates}
\label{sec:find-darb-coord}

A natural problem of the theory of Hamiltonian operators is: \emph{given a
  Hamiltonian operator $A$, find coordinates such that the operator takes the
  form $A^{ij} = \eta^{ij}\partial_x$}, where $(\eta^{ij})$ is a constant
matrix.  The corresponding coordinates are said to be \emph{Darboux
  coordinates} of $A$.  The problem of finding Darboux coordinates for
Hamiltonian operators was considered by many authors so far, like
\cite{AstashovVinogradov:SHOpFT,mokhov87:_hamil,mokhov85:_local_poiss,olver88:_darboux_hamil},
where scalar Hamiltonian operators have been considered.

Darboux coordinates for operators $A_1$ of the type \eqref{eq:27} always exist:
they are flat coordinates for the flat pseudo-Riemannian metric $h_{ij}$. This
means that a point transformation (which is not always easy to find in concrete
examples) is enough to transform an operator  of type \eqref{eq:27} to $A^{ij}
= \eta^{ij}\partial_x$.

A more difficult problem is finding Darboux coordinates for higher-order
Dubrovin--Novikov operators. These operators were introduced in
\cite{DubrovinNovikov:PBHT}. In the third-order case they have the form
\begin{equation}
  A^{ij}=g^{ij}\ddx{3}+b_{k}^{ij}u_{x}^{k}\ddx{2}
  +(c_{k}^{ij}u_{xx}^{k}+c_{km}^{ij}u_{x}^{k}u_{x}^{m})\ddx{}
  +d_{k}^{ij}u_{xxx}^{k}+d_{km}^{ij}u_{xx}^{k}u_{x}^{m}
  +d_{kmn}^{ij}u_{x}^{k}u_{x}^{m}u_{x}^{n}.
  \label{third}
\end{equation}
where coefficients are functions of $(u^i)$. It can be proved that the
coefficient $ - g_{js}d_{k}^{is}$ transforms like a linear connection that is
symmetric and flat by the Hamiltonian property. The operator \eqref{third} can
be rewritten in flat coordinates of $ - g_{js}d_{k}^{is}$ as in
\eqref{eq:24}. \hl{This means that the operator is completely determined by its
  leading term $(g^{ij})$ using} \eqref{eq:195}.  The \hl{pseudo-Riemannian
  metric} $(g_{ij})$ is in bijective correspondence with certain projective
varieties, see \cite{FPV14,FPV16,FPV17:_system_cl}. In particular, in the case
of $3$ dependent variables $u^1$, $u^2$, $u^3$ the operators are divided in $6$
classes with respect to reciprocal transformations of the following projective
type
\begin{equation}
  \label{eq:28}
  d\tilde{x} = (a^0_0+a^0_iu^i)dx,\qquad d\tilde{t} = dt,\qquad
  \tilde{u}^i = \frac{a^i_0+a^i_ju^j}{a^0_0+a^0_ju^j},
\end{equation}
where $a^i_j$, $a^0_j$, $a^i_0$, $a^0_0$ are constants.
The $6$ projective classes of operators are \label{h-op}
    \begin{gather*}
   g^{(1)}=\begin{pmatrix}
(u^{2})^{2}+c & -u^{1}u^{2}-u^{3} & 2u^{2}
\\ -u^{1}u^{2}-u^{3} & (u^{1})^{2}+c(u^{3})^{2} & -cu^{2}u^{3}-u^{1}
\\ 2u^{2} & -cu^{2}u^{3}-u^{1} & c(u^{2})^{2}+1
\end{pmatrix},
\\
  g^{(2)} = \begin{pmatrix}
    (u^{2})^{2}+1 & -u^{1}u^{2}-u^{3} & 2u^{2} \\
    -u^{1}u^{2}-u^{3} & (u^{1})^{2} & -u^{1} \\
    2u^{2} & -u^{1} & 1
  \end{pmatrix}, \quad
    g^{(3)} = \begin{pmatrix}
      (u^{2})^{2}+1 &  -u^{1}u^{2}&0 \\
      -u^1u^2 & (u^1)^2 & 0 \\
      0 & 0 & 1%
    \end{pmatrix},
\\
  g^{(4)}=   \begin{pmatrix}
      -2u^2  & u^1  &  0
\\
u^1  & 0 &   0
\\
0 & 0  & 1
    \end{pmatrix},
\quad
     g^{(5)}=\begin{pmatrix}
      -2u^2 & u^1 & 1
      \\
      u^1 & 1 &0
      \\
      1 & 0 & 0
    \end{pmatrix},
\quad
g^{(6)} =
    \begin{pmatrix}
      1 & 0 & 0\\ 0 & 1 & 0\\ 0 & 0 & 1
    \end{pmatrix}.
  \end{gather*}
  After \hl{introducing} potential coordinates $u^i = b^i_x$ the
  operator~\eqref{eq:24} takes the form
\begin{equation}
  \label{eq:30}
  A^{ij} = - \left(g^{ij}\ddx{} + c^{ij}_k b^k_{xx}\right),
\end{equation}
where coefficients are functions of $(b^i_x)$.  It can be proved
\cite{PV15,FPV17:_system_cl} that each of the above metrics can be factorized
as $g_{ij}=\phi_{\alpha\beta}\psi^\alpha_i\psi^\beta_j$, where
$(\phi_{\alpha\beta})$ is a constant non-degenerate symmetric matrix and
$\psi^\alpha_i$ are linear functions of the field variables:
$\psi^\alpha_i=\psi^\alpha_{im}u^m + \omega^\alpha_i$, where
$\psi^\alpha_{im}=-\psi^\alpha_{mi}$. In \cite{FPV17:_system_cl} we proved that
$n$ Casimirs exist for every operator in the form~\eqref{eq:30}. More
precisely, we proved that the functions
$C^\alpha = \left(\frac{1}{2}\psi^\alpha_{mk}b^k_x+\omega^\alpha_m\right)b^m$
satisfy
\begin{equation}
  \label{eq:31}
  A^{ij}\fd{C^\alpha}{b^j} = 0.
\end{equation}
The above Casimirs turn out to be Darboux coordinates for $g^{(4)}$ and
$g^{(5)}$.  In order to prove that, we use the change of coordinates formula
(see, e.g., \cite{mokhov87:_hamil,olver88:_darboux_hamil})
\begin{equation}
  \label{eq:33}
  \tilde{A} = \ell_C \circ A \circ \ell^*_C,
\end{equation}
where $C$ is the Casimir vector function. The proof that $\tilde{A}$ will be of
Darboux type can be done by \cde. We refer to the program file
\texttt{casimir}. We start by
\begin{rlispverb}
indep_var:={x};
dep_var:={b1,b2,b3};
odd_var:={p1,p2,p3};
total_order:=6;
\end{rlispverb}
and call \texttt{cde}, then we load the metric
\begin{rlispverb}
g_5:=mat(( - 2*b2_x,b1_x,1),(b1_x,1,0),(1,0,0));
gl3:=g_5;
gu3:=gl3**(-1);
\end{rlispverb}
and define the operator \texttt{c\_hi\_con(i,j)} with the values of the
expression $c^{ij}_k b^k_{xx}$ (see the source file) and introduce the operator
$A$ as
\begin{rlispverb}
mk_cdiffop(a,1,{3},3);
for all i,j,psi let a(i,j,psi) =
 - (gu3(i,j)*td(psi,x)+c_hi_con(i,j)*psi);
\end{rlispverb}
The Casimirs \hl{of the operator $A$ are:}
\begin{rlispverb}
operator casimir;
casimir(1):=b1;
casimir(2):=b2;
casimir(3):=b3 + b1_x*b2;
\end{rlispverb}
We linearize the vector function \texttt{casimir} and define its adjoint:
\begin{rlispverb}
f_dar:=for i:=1:ncomp collect casimir(i);
ell_function(f_dar,ldar);
adjoint_cdiffop(ldar,ldar_star);
\end{rlispverb}
The formula~\eqref{eq:33} is easily implemented:
\begin{rlispverb}
mk_cdiffop(ta,1,{3},3);
for all i,j,psi let ta(i,j,psi)=
 for k:=1:ncomp sum for h:=1:ncomp sum
   ldar(i,k,a(k,h,ldar_star(h,j,psi)));
\end{rlispverb}
The result can be converted into a superfunction for a better presentation
\begin{rlispverb}
conv_cdiff2superfun(ta,ta_sf);
\end{rlispverb}
we have
\begin{rlispverb}
ta_sf(1);
 - p3_x
ta_sf(2);
 - p2_x
ta_sf(3);
 - p1_x
\end{rlispverb}
that confirms that the above choice of Casimirs is a set of Darboux coordinates
for $A$. In particular,
\begin{equation}
  \label{eq:34}
  \tilde{A} =
  \begin{pmatrix}
    0 & 0 & -1 \\ 0 & -1 & 0 \\ -1 & 0 & 0
  \end{pmatrix}\partial_x .
\end{equation}
Note that this is a proper differential substitution, i.e., a
transformation that depends on derivatives of the dependent variables.

\begin{problem}
  \hl{Proving that the nonlocal Casimirs are Darboux coordinates for $g^{(4)}$
    is not a problem}
  \cite{kalayci98:_alter_hamil_wdvv,kalayci97:_bi_hamil_wdvv}. \hl{However, the
    author did not manage to prove a similar result for the metrics $g^{(1)}$,
    $g^{(2)}$, $g^{(3)}$ and the metric of the third-order operator in
    subsection}~\ref{sec:warmup:-large-scale}. \hl{Note that the
    equation}~\eqref{eq:33}, \hl{where $\tilde{A}=\eta^{ij}\partial_x$, $\eta^{ij}$ is
    a constant matrix and $C$ is an unknown vector function, is non-linear
      with respect to $C$; the resulting system of equations on $C$ can be
      difficult to solve,} even by using a simplified ansatz. \hl{The
      interested Reader might wish to consider this as a possible research
      problem.}
\end{problem}

\subsection{Compatibility of third-order operators}

\hl{A compatible pair $A_1$, $A_2$ of Hamiltonian operators immediately leads
  to new integrable systems. The standard way to construct them is to solve the
  equation for Casimirs of $A_1$:}
  \begin{equation}
    \label{eq:26}
    A_1^{ij}\fd{C}{u^j} = 0,
  \end{equation}
  \hl{where $C$ is an unknown function of $(u^k)$.  If there are at least $n$
    independent solutions of the above equations, they can be used as new
    coordinates $(\tilde{u}^i)$. In most cases the new coordinates are
    \emph{not} Casimirs of $A_2$, and define an integrable hierarchy through
    Magri's recursion}~\eqref{eq:10}.

  \hl{For this reason, a}n interesting question that can
  be posed is: if we consider all possible pairs of homogeneous third-order
  Hamiltonian operators $A_1$, $A_2$, both from our list on page \pageref{h-op}
  (in the form~\eqref{eq:24}), will there be compatible pairs? The answer is:
  yes, but only in a rather trivial sense. The details are in the following
  table:
\begin{center}
    \begin{tabular}{ | l | l | l | l | l | l | l |}
    \hline
    & $g^{(1)}$ & $g^{(2)}$ & $g^{(3)}$ & $g^{(4)}$ & $g^{(5)}$ & $g^{(6)}$
      \\ \hline
    $g^{(1)}$ & y$^*$ & n & n & n & n & n \\ \hline
    $g^{(2)}$ & n & y & n & n & n & n \\ \hline
    $g^{(3)}$ & n & n & y & y & n & n \\ \hline
    $g^{(4)}$ & n & n & y & y & n & n \\ \hline
    $g^{(5)}$ & n & n & n & n & y & n \\ \hline
    $g^{(6)}$ & n & n & n & n & n & y \\ \hline
    \end{tabular}
  \end{center}
The notation is: $n$ for non compatible, $y$ for compatible, $y^*$ for
compatible under additional conditions. More particularly,
the operators coming from $g^{(3)}$ and $g^{(4)}$ are indeed the direct sum of
a $2\times 2$ block and the one-dimensional operator $\partial_x^3$, and the
$2\times 2$ blocks are known to be compatible~\cite{FPV14}.
Operators $A_1$ and $A_2$ coming from $g^{(1)}$ are compatible if and only if
the value of the constants $c$ in both operators is the same; so, we have no
new compatible pairs besides the known ones.

The code for the above computation is rather simple (file
\texttt{compat3rd}). We define an operator by
\begin{rlispverb}
mk_cdiffop(aa1,1,{ncomp},ncomp);
for all i,j,psi let aa1(i,j,psi) =
td(
  gu3_1(i,j)*td(psi,x,2) + c_hi_con_1(i,j)*td(psi,x)
    ,x
);
\end{rlispverb}
and another operator \texttt{aa2} with a different metric \texttt{gu3\_2(i,j)},
and then convert both to superfunctions and bivectors:
\begin{rlispverb}
conv_cdiff2superfun(aa1,aa1_sf);
conv_cdiff2superfun(aa2,aa2_sf);
conv_genfun2biv(aa1_sf,biv_aa1);
conv_genfun2biv(aa2_sf,biv_aa2);
\end{rlispverb}
Finally, we compute the Schouten bracket and require that the three-vector is a
total divergence:
\begin{rlispverb}
schouten_bracket(biv_aa1,biv_aa2,th12);
eth12:=euler_df(th12(1));
templ:=for each el in eth12 join el;
sb_coeff:=splitext_list(templ);
sb_num_coeff:=for each el in sb_coeff collect num el;
sb_allcoeff:=splitvars_list(sb_num_coeff,all_parametric_der);
\end{rlispverb}
The list \texttt{eth12} contains all variational derivatives with respect to
even and odd coordinates of the Schouten bracket, the list
\texttt{sb\_coeff} contains all coefficients of odd variable expressions, the
list \texttt{sb\_num\_coeff} contains the numerators of the previous
expressions, and the list \texttt{sb\_allcoeff} contains the coefficients of
monomials of even variables. It is easy to deduce the above results.

Despite the fact that the result is negative, it is new and the above
computational scheme can be used for more interesting computations. \hl{The
  above computation would clearly be very hard with pen and paper.}

\begin{problem}
  \hl{The above computation is not a well-posed problem under the geometric
    viewpoint. Indeed, if $A$ and $B$ are two third-order Hamiltonian operators
    of the type}~\eqref{eq:24} \hl{then only one of them (say $A$) can be
    brought to one of the forms}~\eqref{h-op}; \hl{the other will be again of
    the form}~\eqref{eq:24} \hl{but the leading term will not necessarily be in
    the list}~\eqref{h-op}.

  \hl{It is an open problem to check if there are any compatible pairs $A$, $B$,
  where $A$ is one of the operators in the above list and $B$ is another
  operator from the same list after a transformation of type}~\eqref{eq:28}.
\end{problem}

\subsection{Finding compatible operators}

In this subsection we will address the following problem: given a Hamiltonian
operator $A$, find all Hamiltonian operators $B$ that are compatible with
$A$. \hl{As it is known (see also the beginning of the previous subsection),
  compatible pairs of Hamiltonian operators yield integrable systems}
\cite{Magri:SMInHEq}. Usually, the answer is provided for specified forms of
the unknown operator $B$.

We \hl{consider} the above problem in the following formulation: given a
third-order local homogeneous Hamiltonian operator $R$ find all first-order
homogeneous Hamiltonian operators $P$ that are compatible with $R$: $[P,R]=0$.
This problem was completely solved in the case $n=2$ in
\cite{LSV:bi_hamil_kdv}.  In this case there is an affine classification of
third-order homogeneous Hamiltonian operators \cite{FPV14}: they are the three
operators $R_1$, $R_2$, $R_3$ respectively determined by the three
pseudo-Riemannian metrics
\begin{displaymath}
    g^{(1)}=\left(
    \begin{array}{cc}
      1 & 0 \\
      0 & 1
    \end{array}
  \right) ~~
  g^{(2)}=\left(
    \begin{array}{cc}
      -2u^{2} & u^{1} \\
      u^{1} & 0
    \end{array}
  \right), ~~
  g^{(3)}=\left(
    \begin{array}{cc}
      (u^{2})^{2}+1 &-u^{1}u^{2} \\-u^{1}u^{2} & (u^{1})^{2}
    \end{array}
  \right).
\end{displaymath}
One of the results from \cite{LSV:bi_hamil_kdv} is that $P_1$ is  a Hamiltonian
operator compatible with $R_3$ if and only if
\begin{subequations}\label{eq:154}
  \begin{align}
    &g^{11}_1=c_1 u^1 + c_2 u^2+c_3,
    \\
    &g^{12}_1=c_4 u^1 - \frac{c_2}{2u^1} + \frac{c_3 u^2}{ u^1}+ \frac{c_2
      (u^2)^2}{ 2u^1},
    \\
    &g^{22}_1=2 c_4 u^2 + \frac{c_1}{u^1} + \frac{c_5 u^2}{ u^1}- \frac{c_1
      (u^2)^2}{ u^1} +c_6,
  \end{align}
\end{subequations}
where $g_1^{ij}$ are the coefficients of the metric of the first-order operator
$P_1$ (of the form~\eqref{eq:27}) together with the algebraic conditions
\begin{equation}\label{eq:183}
  c_2 c_5 + 2 c_1 c_3 = 0, \quad c_2 c_6 - 2 c_3 c_4 = 0,
  \quad c_1 c_6 + c_4 c_5 = 0.
\end{equation}
Let us set up the computation in \cde; we describe the program file
\texttt{compat13}. After initialization, we define an operator
\texttt{hu1\_op} that contains the leading term of the unknown operator $P_1$:
\texttt{hu1\_op(i,j):=hu1\_ij} where \texttt{hu1\_ij} depends on dependent
variables only. We also define an operator \texttt{gamma\_hi} in such a way
that
\begin{enumerate}
\item \texttt{gamma\_hi(i,j,k):=gamma\_hi\_ijk} for $i<j$
\item \texttt{gamma\_hi(i,i,k):=(1/2)*df(hu1\_op(i,i),part(dep\_var,k))}
\item \texttt{gamma\_hi(j,i,k):= - gamma\_hi(i,j,k)
    + df(hu1\_op(i,j),part(dep\_var,k))} for $i>j$
\end{enumerate}
Indeed, we can use the linear part of the Hamiltonian properties for an
operator of the form~\eqref{eq:27} as given in~\cite{dubrovin98:_flat_froben}
in order to reduce the number of unknowns: they are
\begin{equation}
  \label{eq:35}
  \Gamma^{ij}_k + \Gamma^{ji}_k = \partial_k g^{ij}_1
\end{equation}
Then, we introduce an operator \texttt{gamma\_hi\_con} such that
\texttt{gamma\_hi\_con(i,j)} contains the expression $\Gamma^{ij}_ku^k_x$.
Now, we define the first-order operator
\begin{rlispverb}
mk_cdiffop(aa1,1,{2},2);
for all i,j,psi let aa1(i,j,psi)=
  hu1_op(i,j)*td(psi,x)+gamma_hi_con(i,j)*psi;
\end{rlispverb}
(here \texttt{aa1} stands for $P_1$) and from the metric
\begin{rlispverb}
matrix g2_3(2,2);
g2_3(1,1):=b2**2 + 1;
g2_3(1,2):= - b1*b2;
g2_3(2,1):=g2_3(1,2);
g2_3(2,2):=b1**2;
gl3:=g2_3;
\end{rlispverb}
construct the third-order operator
\begin{rlispverb}
mk_cdiffop(aa2,1,{2},2);
for all i,j,psi let aa2(i,j,psi) =
td(
gu3(i,j)*td(psi,x,2)+c_hi_con(i,j)*td(psi,x)
,x);
\end{rlispverb}
(here \texttt{aa2} stands for $R_3$). After converting the operators into
bivectors
\begin{rlispverb}
conv_cdiff2superfun(aa1,sym1);
conv_cdiff2superfun(aa2,sym2);
conv_genfun2biv(sym1,biv1);
conv_genfun2biv(sym2,biv2);
\end{rlispverb}
it can be easily checked that
$[R_3,R_3]=0$:
\begin{rlispverb}
schouten_bracket(biv2,biv2,sb22);
euler_df(sb22(1));
\end{rlispverb}
The compatibility equation $[P_1,R_3]=0$ is found by
\begin{rlispverb}
schouten_bracket(biv1,biv2,sb12);
comp12:=euler_df(sb12(1));
\end{rlispverb}

Then, we solve the above equation. To do that, we first define an operator
\texttt{equ} whose values are the $4$ components of the equation, then use a
\cde procedure that takes all coefficients of all monomials of odd coordinates
in the equations and put them in the values \texttt{equ(5)}, \texttt{equ(6)},
\dots of the operator:
\begin{rlispverb}
operator equ;
equ(1):=num first first comp12;
equ(2):=num second first comp12;
equ(3):=num first second comp12;
equ(4):=num second second comp12;
tel:=4;
tel_start:=4;
tel:=splitext_opequ(equ,1,4);
\end{rlispverb}
then, we  set up \hl{the solver CRACK for overdetermined PDEs}
\cite{WB95,Wolf93,WB}
\begin{rlispverb}
unk:=append(unk_hu1,unk_gamma_hi);
system_eq:=for i:=tel_start+1:tel collect equ(i);
load_package crack;
lisp(max_gc_counter:=10000000000);
crack_results:=crack(system_eq,{},unk,
   cde_difflist(all_parametric_der,dep_var));
\end{rlispverb}
the solution is~\eqref{eq:154} (after renaming the constants); we are also able
to determine the Christoffel symbols
$\Gamma^{ij}_k$ in terms of the same constants.  \hl{We should require the
  further condition}
\begin{equation}
  \label{eq:36}
  g^{is}\Gamma^{jk}_s = g^{js}\Gamma^{ik}_s
\end{equation}
in order to obtain the symmetry of $\Gamma^i_{jk}$ and the fact that
$\Gamma^i_{jk}$ are the Christoffel symbols of the Levi-Civita connection of
$g^{ij}_1$. It turns out that the metric in~\eqref{eq:154} and the symbols
$\Gamma^{ij}_k$ satisfy the above equation without further conditions.
Moreover, we must require the flatness of the metric $g_1^{ij}$; that amounts
to the algebraic equations~\eqref{eq:183}.

In~\cite{LSV:bi_hamil_kdv} we also prove that there are pencils of compatible
first-order operators inside the space of solutions of the above
problem. Unpublished computations with \cde \hl{show} that such pencils
continue to exist in the case $n=3$ and $n=4$, thus providing multi-parameter
spaces of Miura-inequivalent integrable systems according with the mechanism
which was first discovered in \cite{olver96:_tri_hamil}. \hl{Such results would
  be very hard to achieve without computer-assisted calculations.}

\begin{problem}\label{sec:find-comp-oper}
  \hl{Find the list of all pencils of compatible first-order operators that are
  compatible with a fixed third-order operator in the case $n=3$ or $n=4$. Use
  the classification of third-order operators provided in} \cite{FPV14,FPV16}.
\end{problem}

\subsection{Computing Lie derivatives}
\label{sec:comp-lie-deriv}

Given a pair of compatible Hamiltonian operators $A_1$, $A_2$, it happens in
many cases that there exists a vector field
\begin{equation}
  \tau = \tau^i\pd{}{u^i},\quad\text{where}\quad
  \tau^i=\tau^i(x^\lambda,u^i_\sigma),\label{eq:37}
\end{equation}
such that $A_2=L_\tau A_1 = [\tau,A_1]$, where the bracket is obviously the
Schouten bracket. This is more than just a curiosity: it is an essential
feature of the perturbative approach to the classification of integrable
systems in $1+1$ dimensions initiated in~\cite{dubrovin98:_bi_hamil}. Indeed,
consider a Hamiltonian deformation of a first-order homogeneous Hamiltonian
operator $P_0$ of the type~\eqref{eq:27}
\begin{equation}
P_\epsilon=P_0+\epsilon P_1 + \epsilon^2 P_2 + \cdots\label{eq:38}
\end{equation}
i.e., a one-parameter family of Hamiltonian operators $P_\epsilon$ where
each of the summands $P_1$, $P_2$, \dots is homogeneous of increasing degree.
Then, it can be proved under some reasonable assumptions (see
\cite{degiovanni05:_poiss,mokhov02:_compat_dubrov_novik_hamil_operat}) that the
deformation is always trivial, in the sense that there exists a vector field
$\tau$ such that $P_\epsilon = L_\tau P_0$; this implies the existence of a
formal diffeomorphism $\phi_\epsilon$ such that
$P_\epsilon = \ell_{\phi_\epsilon}\circ P_0\circ\ell^*_{\phi_\epsilon}$.
See \cite{sergyeyev04:_simpl_way_makin_hamil_system} for a detailed exploration
on Lie derivative and compatibility for Hamiltonian operators.

More generally, any Hamiltonian operator $P_0$ defines a map $d_{P_0} =
[P_0,\cdot]$ which is a differential: $d_{P_0}^2=0$. The Hamiltonian
cohomology, or Lichnerowicz--Poisson cohomology of $d_{P_0}$ measures the
presence of operators which are compatible with $P_0$ and are not the Lie
derivative of $P_0$, so that they are not deformable back to $P_0$.

Another interesting feature of the Lie derivative is that it is useful to find
topological hierarchies. Topological hierarchies are related with the theory of
Gromov-Witten invariants, the theory of singularities and other topics; for
more details, see \cite{dubrovin01:_normal_pdes_froben_gromov_witten}.
In \cite{falqui12:_exact_poiss} it was proved that if two
Hamiltonian operators $A_1$ and $A_2$ of the type~\eqref{eq:38} fulfilling some
further property define a topological hierarchy if and only if they are an
\emph{exact Poisson pencil}. The latter condition is, by definition, the
existence of a vector field $\tau$ such that
\begin{equation}
  \label{eq:39}
  A_2 = L_\tau A_1,\qquad L_\tau A_2 = L_\tau^2 A_1 = 0.
\end{equation}

Hence, for the above reasons (and more) it is natural to pose the question:
given two compatible Hamiltonian operators $A_1$, $A_2$ find a vector field
$\tau$ (if it exists!) such that $A_2 = L_\tau A_1$.

In this paper we will consider two such calculations: one is for the
bi-Hamiltonian pair of the KdV-equation, from \cite{degiovanni05:_poiss}, and
the other is for the bi-Hamiltonian pair of the WDVV equation \cite{D96}
presented as a hydrodynamic-type system \cite{PV17:_remar_lagr}.

Let us start with the KdV equation; the calculation is in the file
\texttt{kdv\_lieder}. After the initialization, we define the two well-known
Hamiltonian operators $A_1$ and $A_2$
\begin{rlispverb}
mk_cdiffop(a1,1,{1},1);
for all psi1 let a1(1,1,psi1)=td(psi1,x);

mk_cdiffop(a2,1,{1},1);
for all psi3 let a2(1,1,psi3)=u_x*psi3 + td(psi3,x,3)
  + 2*u*td(psi3,x);
\end{rlispverb}
and consider the vector field $\tau$ as a degree $1$ variational multivector
\texttt{tau} and superfunction \texttt{tau\_sf}
\begin{rlispverb}
mk_cdiffop(tau,1,{1},1);
for all phi let tau(1,phi) = (- (1/2)*u**2 - (1/2)*u_2x)*phi;

mk_superfun(tau_sf,1,1);
tau_sf(1):= (- (1/2)*u**2 - (1/2)*u_2x)*p;
\end{rlispverb}
Then, we convert the above operators into the corresponding bivectors
\begin{rlispverb}
conv_cdiff2superfun(a1,s1);
conv_cdiff2superfun(a2,s2);

conv_genfun2biv(s1,biv1);
conv_genfun2biv(s2,biv2);
\end{rlispverb}
and we compute the Schouten bracket $[\tau, A_2]$:
\begin{rlispverb}
schouten_bracket(tau_sf,biv1,l_tau_biv1);
l_tau_biv1(1);
\end{rlispverb}
The result is not the same as \texttt{biv2(1)}; however, they coincide in the
variational cohomology, as they should:
\begin{rlispverb}
euler_df(l_tau_biv1(1) - biv3(1));

{{0},{0}};
\end{rlispverb}
the latter being the list of the variational derivatives with respect to even
($u$) and odd ($p$) coordinates.

It is known that the simplest WDVV equation:
$f_{ttt}=f_{xxt}^{2}-f_{xxx}f_{xtt}$ can be rewritten as a hydrodynamic-type
system by introducing new variables $a^{1}=a=f_{xxx}$, $a^{2}=b=f_{xxt}$,
$a^{3}=c=f_{xtt}$, namely:
\begin{equation}
a_{t}=b_{x}, \quad b_{t}=c_{x}, \quad c_{t}=(b^{2}-ac)_{x}.  \label{eq:511}
\end{equation}
It was proved in \cite{FGMN97} that the above system is bi-Hamiltonian, with
(compatible) operators 
\begin{gather}
A_{1}=%
\begin{pmatrix}
-\frac{3}{2}\partial _{x}^{{}} & \frac{1}{2}\partial _{x}^{{}}a & \partial
_{x}^{{}}b \\
\frac{1}{2}a\partial _{x}^{{}} & \frac{1}{2}(\partial _{x}^{{}}b+b\partial
_{x}^{{}}) & \frac{3}{2}c\partial _{x}^{{}}+c_{x} \\
b\partial _{x}^{{}} & \frac{3}{2}\partial _{x}^{{}}c-c_{x} &
(b^{2}-ac)\partial _{x}^{{}}+\partial _{x}^{{}}(b^{2}-ac)%
\end{pmatrix},
\\
A_{2}=%
\begin{pmatrix}
0 & 0 & \partial _{x}^{3} \\
0 & \partial _{x}^{3} & -\partial _{x}^{2}a\partial _{x} \\
\partial _{x}^{3} & -\partial _{x}a\partial _{x}^{2} & \partial
_{x}^{2}b\partial _{x}+\partial _{x}b\partial _{x}^{2}+\partial
_{x}a\partial _{x}a\partial _{x}%
\end{pmatrix}.%
\end{gather}%
We can rewrite the system and the operators in flat coordinates~$u^{k}$ of the
leading term of the operator $A_1$ (as a flat pseudo-Riemannian contravariant
metric):
\begin{equation}
  a=u^1+u^2+u^3,
\quad
  b=-\frac{1}{2}(u^1u^2+u^2u^3+u^3u^1),
  \quad
  c=u^1u^2u^3,
\end{equation}
the operator $A_{1}$ becomes $A_{1}^{ij}=K^{ij}\partial _{x}^{{}}$, where
\begin{equation*}
(K^{ij})=\frac{1}{2}\left(
\begin{array}{ccc}
1 & -1 & -1 \\
-1 & 1 & -1 \\
-1 & -1 & 1%
\end{array}%
\right),
\end{equation*}
and the system~\eqref{eq:511} takes the form
\begin{equation}
  \label{eq:323}
    u^i_t=\frac{1}{2}(u^ju^k-u^iu^j-u^iu^k)_x,
\end{equation}
where $i$, $j$, $k$ are three distinct indices. Note that
flat coordinates $u^i$ of the first operator $A_1$ make the expression of the
second operator $A_2$ much more complicated with respect to the initial
coordinates $a$, $b$, $c$.

It is proved in \cite{PV17:_remar_lagr} that the coordinate expression of
$\tau$ is
\begin{equation}
  \label{eq:40}
  \tau = - K^{in}L_n\frac{\partial}{\partial u^i},\quad\text{where}\quad
  L_{n}=\left( \frac{1}{2}G_{nm}u_{x}^{m}+R_{nm}u_{x}^{m}\right) _{x}-\frac{1}{%
2}L_{nsm}u_{x}^{s}u_{x}^{m}.
\end{equation}
In the above formula we have:
\begin{enumerate}
\item $-\frac{1}{%
2}L_{nsm}u_{x}^{s}u_{x}^{m}$ are $3$ conserved densities of the
system~\eqref{eq:323} with coordinate expressions (note that $L_{ijk}=L_{ikj}$):
\begin{enumerate}
\item when $j\neq 1$ and $k\neq 1$ or when $j=k=1$
\begin{equation}
  \label{eq:186}
  L_{1jk}=\frac{((u^1-u^2)+(u^1-u^3))(u^{a}-u^{b})(u^{c}-u^{d})}
        {2(u^1-u^2)^3(u^1-u^3)^3},
\end{equation}
where $(a,j,b)$, $(c,d,k)$ are triplets of distinct indices with $a<b$, $c<d$;
\item when $(1,j,k)$ are a triplet of distinct indices
\begin{equation}
  \label{eq:187}
  L_{11k}= - \frac{(u^{1}-u^{k})^2+(u^{1}-u^{j})^2}
        {2(u^1-u^{k})^2(u^1-u^{j})^3}.
\end{equation}
\item when $i\neq 1$ the expressions of $L_{ijk}$ are obtained by a cyclic
  permutation of the above expressions.
\end{enumerate}
\item $G_{nm}$ is the leading term of the symplectic operator
  $B_{ij} = - K_{ip}A^{pq}_2K_{qj}$ (see \cite{PV17:_remar_lagr}), and it is a
  pseudo-Riemannian metric; the coordinate expression of the inverse matrix
  $(G^{ij})$ is simpler:
  \begin{equation}\label{eq:42}
  G^{ii}=-\frac{1}{4}(u^i-u^j)(u^i-u^k),\quad
  G^{ij}=-\frac{1}{4}(u^i-u^j)^2.
  \end{equation}
\item $R_{pm}$ is a two-form whose expression is
  \begin{equation}
    \label{eq:41}
    R_{ij}= - \frac{1}{3}\left( \frac{1}{(u^{i}-u^{j})(u^{i}-u^{k})}
      - \frac{1}{(u^{j}-u^{i})(u^{j}-u^{k})}\right),
\end{equation}
where $i$, $j$, $k$ are distinct indices.
\end{enumerate}

Let us describe the computation that leads to directly proving the equality
$L_\tau A_1 = A_2$ (file \texttt{w3c\_lagrep3}). We initialize the jet space
with
\begin{rlispverb}
indep_var:={x};
dep_var:={u1,u2,u3};
odd_var:={p1,p2,p3};
total_order:=10;
ncomp:=length(dep_var);
\end{rlispverb}
We load the third-order
operator $A_2$ as in subsection~\ref{sec:warmup:-large-scale}, then we set up
the change of variables with its Jacobian:
\begin{rlispverb}
a:=u1 + u2 + u3;
a_x:=td(a,x);
b:=-1/2*(u1*u2 + u2*u3 + u3*u1);
b_x:=td(b,x);
c:=u1*u2*u3;
c_x:=td(c,x);
a_uqs:={a,b,c};

matrix jac(ncomp,ncomp);
for i:=1:ncomp do
  for j:=1:ncomp do jac(i,j):=df(part(a_uqs,i),part(dep_var,j));
jacinv:=jac**(-1);
\end{rlispverb}
We can use the following formula to change coordinates to the operator $A_2$:
\begin{equation}
  \label{eq:1611}
  A_2^{ij}(\mathbf{u})=\frac{\partial u^{i}}{\partial a^{n}}
   A^{nm}_2(\mathbf{a})\frac{\partial u^{j}}{\partial a^{m}},
\end{equation}
implemented as
\begin{rlispverb}
mk_cdiffop(taa2,1,{3},3);
for all i,j,psi let taa2(i,j,psi)=
  (for h:=1:ncomp sum
    (for k:=1:ncomp sum jacinv(i,h)*aa2(h,k,jacinv(j,k)*psi))
   );
\end{rlispverb}
Obviously, \texttt{taa2} is the operator $A_2$ in the coordinates $u^i$. The
various tensors $G_{mn}$, $R_{mn}$ and $L_{nmp}$ can be obtained from the
coefficients of the operator  \texttt{taa2} (see the file
\texttt{w3c\_lagrep3}). The operator \texttt{l\_n} can be defined in such a way
that the expression of \texttt{l\_n(n)} will be $L_n$. Then we define the
operator $A_1$ (in coordinates $u^i$)
\begin{rlispverb}
matrix kap(ncomp,ncomp),kapinv(ncomp,ncomp);
kap:=(1/2)*mat((1,-1,-1),(-1,1,-1),(-1,-1,1));
kapinv:=kap**(-1);

mk_cdiffop(taa1,1,{3},3);
for all i,j,psi let taa1(i,j,psi) =
kap(i,j)*td(psi,x);
\end{rlispverb}
and vector field $\tau$, with its conversion to an operator and a superfunction
\begin{rlispverb}
mk_cdiffop(tau,1,{3},1);
for all i,phi let tau(i,1,phi)=
  (for j:=1:ncomp sum - kap(i,j)*l_n(j))*phi;
mk_superfun(tau_sf,1,1);
tau_sf(1):=for i:=1:ncomp sum tau(i,1,part(odd_var,i));
\end{rlispverb}
After the conversion of the Hamiltonian operators to bivectors:
\begin{rlispverb}
conv_cdiff2superfun(taa1,taa1_sf);
conv_cdiff2superfun(taa2,taa2_sf);
conv_genfun2biv(taa1_sf,biv1);
conv_genfun2biv(taa2_sf,biv2);
\end{rlispverb}
we can compute the Lie derivative $L_\tau A_1$ as the Schouten bracket
$[\tau,A_1]$
\begin{rlispverb}
schouten_bracket(tau_sf,biv1,l_tau_biv1);
\end{rlispverb}
and check that the result coincides with \texttt{biv2} in the variational
cohomology:
\begin{rlispverb}
euler_df(l_tau_biv1(1) - biv2(1));
\end{rlispverb}
Beware: the computation takes 18 hours and 8GB RAM on a (not too fast) compute
server.

\begin{problem}
  \hl{Repeat the above computation for the bi-Hamiltonian pair of the WDVV
    equation in the case $N=4$} \cite{PV15}. \hl{Can the computation be
    successfully carried out for the oriented associativity equation}
    \cite{casati19:_hamil}?
\end{problem}

\begin{remark}
  A few simple examples of computations of Schouten bracket can be found in the
  \cde manual (which is part of the official Reduce manual).

  \hl{We did not show any calculation of Schouten bracket between operators in
    more than one independent variable due to space constraints.} \cde \hl{has
    this capability, the interested Reader can find one simple example (provided
    by Casati) in the} \cde \hl{manual}.  More non-trivial examples, with
  applications to Hamiltonian and bi-Hamiltonian cohomology, are computed (also
  by \cde) in the recent paper \cite{casati17:_higher_poiss}.
\end{remark}

\section{Conclusions}

In this paper we deliberately did not deal with nonlocal operators, i.e.,
operators which contain expressions with $\partial_x^{-1}$. At the moment, the
only publicly available software for computations of Schouten bracket between
nonlocal operators is the Maple package \texttt{JET}, but it has some
limitations: \hl{indeed, it cannot always simplify expressions that contain
  $\partial_x^{-1}$. Recently, an algorithm for bringing such expression to a
  canonical form has been introduced} \cite{CLV19}, \hl{and we plan to
  implement it as an additional module of} \cde \hl{in the near future.} \hl{An
  extension of the odd variable formalism for the Schouten bracket of nonlocal
  operators is in progress} \cite{krasilshchik18:_nonloc_jacob}.

An interesting perspective of application of \cde could be in theoretical
physics, wherever supermanifolds play a role. Indeed, \cde can do \hl{every
calculation on jets of superbundles} (see
\cite{IgoninVerbovetskyVitolo:VMBGJS,KerstenKrasilshchikVerbovetsky:HOpC,KV11}).
\hl{This means, for example, Euler--Lagrange expressions of superfunctions,
symmetries and conserved quantities of super-differential equations, etc..}
Hence, it could be of help in complex computations in supersymmetric mechanics
and field theories.

\subsection*{Acknowledgments} I would like to thank P. Kersten and
A. Norman for their invaluable \reduce support. Thanks are due to M. Casati,
B.A. Dubrovin, E.V. Ferapontov, A. Fordy, J.S. Krasil'shchik, M.V. Pavlov,
A. Sergyeyev, A.M. Verbovetsky for scientific discussions and motivation to
write and improve the software.

Finally, I would like to thank A. Falconieri, the system administrator of the
workstation \texttt{sophus2} of the Department of Mathematics and Physics of
the Universit\`a del Salento.

This research has been funded by the Dept. of Mathematics and Physics
  ``E. De Giorgi'' of the Universit\`a del Salento, Istituto Naz. di Fisica
  Nucleare IS-CSN4 \emph{Mathematical Methods of Nonlinear Physics}, GNFM of
  Istituto Nazionale di Alta Matematica.


\end{document}